\documentstyle[12pt]{article}
\def\lromn#1{\uppercase\expandafter{\romannumeral#1}}

\def\blist{\begin{list}{\setlength{\rightmargin}{\leftmargin}}}
\def\elist{\end{list}}

\def\CMP{{\sl Commun.\ Math.\ Phys.\ }}

\def\JMP{{\sl J.\ Math.\ Phys.\ }}

\def\PL{{\sl Phys.\ Lett.\ }}

\def\PRep{{\sl Phys.\ Rep.\ }} 
\def\PR{{\sl Phys.\ Rev. }}

\def\PRL{{\sl Phys.\ Rev.\ Lett.\ }}

\begin{document}

\begin{titlepage}

\begin{flushright}

\begin{large}
TU/94/468\\
September 1994
\end{large}
\end{flushright}

\vspace{12pt}

\begin{center}
\begin{Large}

\renewcommand{\thefootnote}{\fnsymbol{footnote}}
\bf{
Time Evolution of Pure Quantum State and Emergence of A Nearly Thermal State
}\footnote[1]
{Work supported in part by the Grant-in-Aid for Science Research from
the Ministry of \\ \hspace*{0.6cm} Education, Science and Culture of
Japan No. 06640367}

\end{Large}

\vspace{36pt}

\begin{large}
\renewcommand{\thefootnote}{\fnsymbol{footnote}}
M.Yoshimura\footnote[2]
{E-mail address: YOSHIM@tuhep.phys.tohoku.ac.jp\@.}\\
Department of Physics, Tohoku University\\
Sendai 980 Japan\\
\end{large}

\vspace{54pt}

{\bf Abstract}\\
\end{center}

Dynamical evolution of the quantum ground state (vacuum) is analyzed
for time variant harmonic oscillators characterized by
asymptotically constant frequency. The oscillatory density matrix
in the asymptotic future is uniquely determined by a constant number
of produced particles, independent of other details of
transient behavior at intermediate times.
Time average over one oscillation period yields a classical,
in some cases even an almost thermal behavior.
In an analytically soluble model
the created particle number obeys the Planck distribution in a parameter
limit.
This suggests a new way of understanding the Gibbons-Hawking temperature in
the de Sitter spacetime.

\end{titlepage}

There are a number of situations in which a precise knowledge
of evolved quantum state in the simple system of time variant harmonic
oscillators is expected to deepen our understanding of fundamental problems
in physics. A class of such problems of great importance is emergence of
classical, in some cases even thermal behavior, resulting from time evolution
of pure quantum ground state in a nontrivial spacetime.
The central issue here is the behavior of quantum field in the curved
spacetime, the quantum field being
an infinite collection of harmonic oscillators with
different frequencies, interacting each other in some prescribed way.
One may mention as examples of such problems the
thermal Hawking radiation from formation of black holes \cite{hw75}, and
quantum dissipation and fluctuation in the inflationary de Sitter universe
 \cite{infldens}.
Although differences in indivisual cases are important, there are some common
features that deserve special attention.
In the classical approximation to gravity the spacetime in the examples listed
is highly curved or is endowed with event horizon.
In an idealization that extracts essential features of the underlying physics,
one replaces the background spacetime by a simpler time evolving space.
For instance, the black hole formation may be modeled by a space bounded
by moving mirror \cite{witt75} that acts as the source of red-shift
just as in the gravitational potential of the black hole.
Investigation of the behavior of quantum fields in such a moving mirror
spacetime has led to Hawking radiation by taking the mirror trajectory
corresponding to the black hole formation.
This picture is ragarded as an essence of the thermal behavior
of the black hole spacetime.

In this paper we would like to lay down foundation of these and other
related problems by
solving the most basic and the simplest problem of this nature:
time evolution of the quantum ground state for time variant harmonic
oscillator.
Time variation of harmonic oscillator frequency is taken arbitrary here,
provided that the initial and the final frequencies approach constant values.
The same problem has undoubtedly been examined many times
in many areas of physics,
but to our knowledge the final quantum wave function or
its density matrix has never been worked out with sufficient generality.
For instance, Parker and
Zeldovich \cite{parker-zeldovich} only discussed rate of
particle production in an expanding universe.
Our construction of the state vector clarifies more details;
we shall be able to write
down the quantum density matrix itself in terms of a few parameters in such
a way that detailed knowledge of time evolution at intermediate times is
not needed. Only with such an apparatus one may discuss the delicate issue of
transition from the quantum to the classical behavior
with some kind of coarse graining. We consider
in this paper the most straightforward coarse graining: time average over
one oscillation period.
This average is also related to a coarse graining in the phase space variable
of the Wigner function.
Even this simple time average yields a new insight
into the fundamental problem set forth; how the thermal behavior emerges
from a quantum state. Our emphasis is on dynamical time evolution,
rather than on
geometry such as the event horizon discussed in many of the recent literature
\cite{hw75},\cite{witt75}, \cite{geometric entropy}.

To illustrate our method of analysis we also briefly sketch an exactly soluble
model of time variation. When suitably interpretated in a parameter limit,
this model describes
evolution of the quantum ground state in the de Sitter spacetime
relevant to the inflationary universe.
The final state in this limit is a nearly thermal one, nearly because
the distribution of produced particles is exactly Planckian, but the time
averaged density matrix has a deviation from the thermal one.
Our model suggests a
new way of understanding the Gibbons-Hawking temperature \cite{gh},
but further work is needed to conclude more definitely.

Quite apart from our motivation described above,
our explicit solution to the problem of time variant harmonic
oscillator is expected to be useful in
solving other fundamental and practical problems in many areas of physics.
With this in mind we stress
the technical aspect of our approach in the main body of the paper.
In the end of this paper we come back to physical applications of
the mathematical analysis.

Consider the generalized harmonic oscillator given by the following
Hamiltonian:
\begin{eqnarray}
H = \frac{1}{2}\, p^{2} + \frac{1}{2}\, \omega ^{2}(t)\,q^{2} \,.
\end{eqnarray}
The variant oscillator frequency $\omega (t)$ may have an arbitrary time
dependence provided that it has a definite limit in both the asymptotic
past and the future:
\( \:
\omega (t) \rightarrow \omega _{0} \hspace{0.3cm} {\rm as \; }
t \rightarrow -\,\infty \,, \hspace{0.5cm}
\omega (t) \rightarrow \omega  \hspace{0.3cm} {\rm as \;}
t \rightarrow \infty \,.
\: \)
For a general $\omega (t)$
the wave function at any finite time is difficult to write down in a useful
form. But as we shall explicitly construct in the following, the asymptotic
form of the transition amplitude can be worked out with only a few parameters
by using the path integral formula. We shall slightly
extend the original Gelfand-Yaglom method \cite{gelfand-yaglom}
by simultaneously solving two boundary
value problems, with the Cauchy data given either at the given initial time
$t_{I}$ or at the given final time $t_{F}$.

The starting point of this construction is the expression for the transition
amplitude when the discretized time slicing with a step $\epsilon
= (t_{F} - t_{I})/(N+1) $
is employed:
\begin{eqnarray}
&&
\langle \:q_{F} \;; t_{F} \:| \:q_{I}\;;t_{I} \:\rangle =
\lim_{N \rightarrow \infty }\,\int\,\frac{1}{\sqrt{2\pi i\epsilon }}\,
\prod _{i}^{N}\,\frac{dq_{i}}{\sqrt{2\pi i\epsilon }}
\nonumber \\
&& \hspace*{0.5cm}
\exp [\:
\frac{i}{2 \epsilon }\,\sum_{i\,, j = 1}^{N}\,q_{i}K^{ij}q_{j} -
\frac{i}{\epsilon }\,\vec{c}\cdot\vec{q} +
\frac{i}{2 \epsilon }\,(K^{F F}q_{F}^{2} + K^{I I}q_{I}^{2})
\:] \,.
\end{eqnarray}
The matrix $K^{ij}$ has nonvanishing elements only along the diagonal and
the next off-diagonal lines:
\( \:
K^{i\,  i} = 2 - \epsilon ^{2}\omega _{i}^{2} \,, \hspace{0.5cm}
K^{i \,, i+1} = K^{i+1 \,, i} = -\,1 \,.
\: \)
Furthermore
\( \:
K^{F F} = 2 - \epsilon ^{2} \omega_{N+1} ^{2} \,, \hspace{0.5cm}
K^{I I} = 2 - \epsilon ^{2} \omega_{0} ^{2} \,, \hspace{0.5cm}
\vec{c} = (\:q_{I}\,, 0\,, \cdots \,, 0 \,, q_{F} \:) \,.
\: \)
Integration with respect to the oscillator coordinates $q_{i}$ yields the
transition amplitude in terms of the inverse matrix $K^{-\,1}$:
\( \:
K^{-1}_{i\,j} = \frac{D_{j-1}^{u}D_{N-i}^{d}}{D_{N}} \,,
\hspace{0.3cm} {\rm for} \hspace{0.3cm} i \geq j \,.
\: \)
$D^{u \:, \:d}_{i}$ is the determinant made out of a part of the matrix $K$ of
dimension $i$, in the upper left (u) or in the lower right (d) corner.
They obey recursion relation with smaller dimensions:
\( \:
D_{i} = K^{i\, i}D_{i-1} - D_{i-2}
\: \)
for both $ D^{u \:, \:d}_{i}$.

In the continuum limit of
\( \:
\epsilon \rightarrow 0
\: \)
the recursion relation combined with the end point value of $D^{u\:, \:d}$
leads to the Gelfand-Yaglom equation with a definite boundary condition.
Thus with $\tilde{D}_{i} \equiv \epsilon D_{i}$
\begin{eqnarray}
(\frac{d^{2}}{dt^{2}} + \omega ^{2}(t))\,\tilde{D}(t) = 0 \,.
\label{gelfand-yaglom eq}
\end{eqnarray}
The boundary conditions are
\begin{eqnarray}
&&
\tilde{D}^{u}(t_{I}) = 0 \,, \hspace{0.5cm}
\frac{d}{dt}\tilde{D}^{u}(t_{I}) = 1 \,, \\
&&
\tilde{D}^{d}(t_{F}) = 0 \,, \hspace{0.5cm}
\frac{d}{dt}\tilde{D}^{d}(t_{F}) = -\,1 \,.
\end{eqnarray}
The continuum limit of the transition amplitude is then given by
\begin{eqnarray}
&&
\langle \:q_{F} \;; t_{F} \:| \:q_{I}\;;t_{I} \:\rangle =
\left( \frac{C}{2\pi i} \right)^{1/2} \,
\exp [\:\frac{i}{2}\,
\,(Aq_{I}^{2} + Bq_{F}^{2} - 2Cq_{I}q_{F}) \:] \,, \nonumber \\
&& \hspace*{-0.5cm}
A = -\,\frac{d}{dt_{I}}\,\ln \tilde{D}^{d}\,(t_{I}) \,, \hspace{0.5cm}
B = -\,\frac{d}{dt_{F}}\,\ln \tilde{D}^{u}\,(t_{F})
\,, \hspace{0.5cm}
C = \frac{1}{\tilde{D}^{d}(t_{I})} = \frac{1}{\tilde{D}^{u}(t_{F})}
\,. \label{transition amp}
\end{eqnarray}
In the rest of this paper we shall be concerned with evolution of the
ground state. The evolved ground state at time $t_{F}$ is obtained,
either by multiplying to Eq.\ref{transition amp}
the initial ground state wave function and integrating with respest to $q_{I}$,
or by directly taking the analytically continued limit of
\( \:
it_{I} \longrightarrow -\,\infty
\: \)
in Eq.\ref{transition amp}.

One may consider an analogue potential model of zero energy under the potential
\( \:
-\,\omega ^{2}(x)
\: \)
by replacing the time $t$ by a space coordinate $x$.
In the far left ($x < 0 $) region the potential approaches
$-\,\omega _{0}^{2}\,$, while in the far right $-\,\omega ^{2}$.
The Gelfand-Yaglom boundary condition is very specific, for instance,
$\tilde{D}^{u}(x)$ must have the form,
\( \:
\sin (\,\omega _{0}(x - t_{I})\,)
/\omega _{0}
\: \)
in the far left near $x = t_{I}$.

In solving the Gelfand-Yaglom problem it would be of great value
if one can decouple the detailed transient behavior at any finite time
and concentrates on asymptotic past and future regions.
This approach is useful unless one asks minute details at intermediate
times. We thus parametrize the asymptotic form of solution respecting
the specified boundary condition. Let us discuss
$\tilde{D}^{u}(t)$ in some detail. It is real by definition
and is given in terms of
two linearly independent solutions, $\psi \; {\rm and \;} \psi ^{*} \;$
to the differential equation Eq.\ref{gelfand-yaglom eq}:
\begin{eqnarray}
&& \hspace*{-1cm}
\tilde{D}^{u}(t)
\equiv \psi  + \psi ^{*}
\rightarrow \frac{1}{2i\omega _{0}}\,(\,
e^{i\omega _{0}(t - t_{I})} - e^{-\,i\omega _{0}(t - t_{I})}\,)
\,, \hspace{0.5cm}
{\rm at \; t\;} \sim t_{I} \sim -\,\infty \,,
\\
&& \hspace*{-1cm}
\psi (t) \:\rightarrow\: a_{1}e^{i\omega t + i\delta _{1}} +
a_{2}e^{-\,i\omega t - i\delta _{2} } \,, \hspace{0.5cm}
{\rm at \; t\;} \sim t_{F} \sim \infty
\,, \label{asymptotic sol}
\end{eqnarray}
where $a_{i}$ is defined as both real and positive, and $\delta _{i}$ are
real phases. Of course, with given frequency variation $\omega (t)$
these parameters
$a_{i} \,, \; \delta _{i}$ are determined by solving the differential equation.
Without resort to explicit form of the solution
we shall be able to derive certain useful results by considering
general properties of these quantities.
We immediately note that
\( \:
a_{1}^{2} - a_{2}^{2} = \frac{1}{4\omega _{0}\omega } \,,
\: \)
verified from constancy of the Wronskian,
and the phases $\delta _{i}$ depend on the initial time like
\( \:
\delta _{1} \sim  -\,\omega _{0}t_{I} \,, \hspace{0.3cm}
\delta _{2} \sim  \omega _{0}t_{I} \,.
\: \)

Similarly one derives the asymptotic form of $\tilde{D}^{d}$.
Starting from \\
\( \:
\tilde{D}^{d}(t)
\sim \frac{1}{2i\omega }\,e^{i\omega (t_{F} - t)} + ({\rm c.c.})
\: \)
near $t_{F}$, one defines at $t \sim t_{I} \sim -\,\infty $
\begin{eqnarray}
\tilde{D}^{d}(t) \:\longrightarrow  \: \tilde{a}_{1}e^{-\,i\omega _{0}t - i
\tilde{\delta} _{1}} + \tilde{a}_{2}e^{i\omega _{0}t + i\tilde{\delta} _{2}}
+ ({\rm c.c.}) \,.
\end{eqnarray}
In terms of these parameters one finds that the quantities
in the transition amplitude
\( \:
\langle \:q_{F} \;; t_{F} \:| \:q_{I}\;;t_{I} \:\rangle
\: \)
in Eq.\ref{transition amp} is given by
\begin{eqnarray}
&&
A = \omega _{0} \cot (\omega _{0}t_{I} + \varphi _{I}) \,,
\hspace{0.5cm}
B =
\omega \cot (\omega t_{F} + \varphi _{F}) \,, \nonumber \\
&& \hspace*{-2cm}
C = [\:2|\:a_{1}e^{i\delta _{1}} + a_{2}e^{i\delta _{2}}\:|\,
\sin (\omega  t_{F} + \varphi _{F}) \:]^{-\,1}
= [\:2|\:\tilde{a}_{1}e^{i\tilde{\delta} _{1}} +
\tilde{a}_{2}e^{i\tilde{\delta} _{2}}\:|\,
\sin (\omega_{0}t_{I} + \varphi _{I}) \:]^{-\,1} \,,
\label{coefficients}
\end{eqnarray}
where $\varphi _{F}$ is independent of $t_{F}$ and
\begin{eqnarray}
-\,\cot \varphi _{F} = \frac{a_{1}\sin \delta _{1} + a_{2}\sin \delta _{2}}
{a_{1}\cos \delta _{1} + a_{2}\cos \delta _{2}} \,.
\end{eqnarray}
A similar relation holds for $\varphi _{I}$ with $a_{i}$ and
$\delta _{i}$ replaced by respective tilde quantities.

One may take the $t_{I} \rightarrow i\infty $ limit in order
to extract the wave function of the evolved vacuum state.
Dependence on the initial time $t_{I}$ in Eq.\ref{coefficients},
besides the explicit one, is through $\delta _{i}$  as already mentioned.
It is convenient to define a quantity $\Lambda $ by
\begin{eqnarray}
a_{1} \equiv \frac{1}{2\sqrt{\omega _{0}\omega }}\,\cosh \Lambda \,,
\hspace{0.5cm}
a_{2} \equiv \frac{1}{2\sqrt{\omega _{0}\omega }}\,\sinh \Lambda \,.
\end{eqnarray}
In the analogue potential model
\( \:
\tanh ^{2}(\Lambda ) = (a_{2}/a_{1})^{2}
\: \)
is  the fraction of left-mover relative to right-mover in the
far right region.
A straightforward computation  shows that the wave function of
the evolved vacuum state is given by
\begin{eqnarray}
&&
\langle \:q_{F}\;t_{F} \:|\:0 \:\rangle  =
(\; {\rm irrelevant \; phase \;factor \;} )\, \times \,
\nonumber \\
&& \hspace*{-1.5cm}
\left( \:\frac{\pi }{\omega }\,(\cosh (2\Lambda) + \sinh (2\Lambda) \cos F) \:
\right)^{-1/4}\,
\exp [\:-\,\frac{\omega }{2}\,q_{F}^{2}\,\frac{1 + i\sinh (2\Lambda) \sin F}
{\cosh (2\Lambda) + \sinh (2\Lambda) \cos F} \:] \,,
\label{wave function}
\end{eqnarray}
where the oscillaroty factor is given by
\begin{eqnarray}
F = 2\omega t_{F} +
(\,{\rm phase \; of \;} a_{1}e^{i\delta _{1} + i\omega _{0}t_{I}} \,)
+ (\,{\rm phase \; of \;} a_{2}e^{i\delta _{2} - i\omega _{0}t_{I}}\,) \,.
\label{phase factor}
\end{eqnarray}
The Wigner transform of the density matrix, namely the Wigner function,
takes a particularly simple form. It is calculated as
\begin{eqnarray}
&& \hspace*{-2cm}
f_{W}(q \,, p \; ; t_{F}) =
2\exp [\:-\,\frac{1}{\omega }\,\cosh (2\Lambda)\,(p^{2} + \omega ^{2}q^{2})
- \frac{1}{\omega }\,\sinh (2\Lambda)\,(p^{2} + \omega ^{2}q^{2})\,
\cos (F + \chi )
\:]
\,, \label{wigner function} \nonumber \\
\end{eqnarray}
where
\( \:
\tan \chi = 2\omega qp/(\omega ^{2}q^{2} - p^{2}) \,.
\: \)

A remarkable feature of the general formula just derived
is that the wave function and the Wigner function depend
on the detailed transient behavior at intermediate times only via the quantity
$\Lambda $. The other phase factor is contained in $F$ Eq.\ref{phase factor}
together with the final time $t_{F}$.
Furthermore the role of the Hamiltonian $H$ is distinct in
Eq.\ref{wigner function}:
all points on the energy surface of fixed $H$
in the phase space ($q\,, p$) run periodically in the
same range of the Wigner function, the exponent varying sinusoidally within
\( \:
-\,2e^{2\Lambda }H/\omega \sim
-\,2e^{-\,2\Lambda }H/\omega
\: \).
Compare this with the value for the quantum ground state,
$-\,2H/\omega $, and that for the thermal state,
\( \:
-\,\frac{2H}{\omega }\,\tanh (\beta \omega /2) \,.
\: \)
Time variation of the potential raises the possibility of particle
production in the evolved state from the initial vacuum, which is indeed
what takes place in this case.
The quantity $\Lambda $ is related to the final number of produced particles
as follows.
By using the standard formula of the creation and the annihilation operators
\( \:
a_{\omega } = -\,i\sqrt{\frac{\omega }{2}}\,q + \frac{1}{\sqrt{2\omega }}\,p
\,, \: \)
one computes, with the wave function Eq.\ref{wave function}, \hspace{0.2cm}
\( \:
\langle \:a_{\omega }^{\dag }a_{\omega } \: \rangle\,(t_{F})
= \sinh ^{2}\Lambda  \,,
\: \)
which is also equal to $4\omega _{0}\omega a_{2}^{2}$.
The produced number of particles is thus independent of time
$t_{F}$ asymptotically:
no oscillation occurs for this measurable quantity.

Rapid oscillation of the Wigner function in the time scale of $1/\omega $
naturally raises the question as to its time averaged
behavior. Certainly one loses important quantum information by the procedure
of time average, but some kind of coarse graining is of prime importance
in statistical interpretation.
The time average of the derived Wigner function Eq.\ref{wigner function}
over one oscillation period is
\begin{eqnarray}
\hspace*{-1cm}
\frac{1}{4\pi }\,\int_{0}^{4\pi }\,dF\,f_{W}(q \,, p \; ; t_{F})
=
2\,\exp [\:-\,\frac{\cosh 2\Lambda }{\omega }\,(p^{2} + \omega ^{2}q^{2})\:]
\,I_{0}[\,\frac{\sinh 2\Lambda }{\omega }\,(p^{2} + \omega ^{2}q^{2})\,] \,,
\label{coarse grained wf}
\end{eqnarray}
where $I_{0}(x)$ is the modified Bessel function. This time-averaged Wigner
function is diagonal in the Hamiltonian basis, and in the large argument
limit it reads as
\begin{eqnarray}
2\,\left( \frac{2\pi }{\omega}\,\sinh (2\Lambda)\,(p^{2} + \omega ^{2}q^{2})
\right)^{-1/2}\,
\exp [\:-\,\frac{1}{\omega}\,\exp (2\Lambda)
\,(p^{2} + \omega ^{2}q^{2}) \:] \,,
\end{eqnarray}
for $ p^{2} + \omega ^{2}q^{2} \gg 1$.
The density matrix $\rho $
corresponding to Eq.\ref{coarse grained wf} does not
obey the relation, $\rho ^{2} = \rho $, required for a quantum state,
due to the average procedure.
The time average over the period can be shown to be equivalent to a coarse
graining over the phase space region of variables ($\:q \,, \;p \:$) of
the Wigner function. Namely, from the $\chi $ dependence of
Eq.\ref{wigner function} one finds the identical coarse grained Wigner
function Eq.\ref{coarse grained wf} by taking the average in the range of
\( \:
|\Delta p| = \omega |\Delta q| =  \sqrt{2H}
\: \)
on the fixed energy ($H$) ellipse.
This exactly corresponds to the classical phase space motion over one
period for the fixed frequency $\omega $.

In physical applications of the mathematical analysis above it often
becomes necessary to extend to a system of many oscillators,
infinitely many in the case of quantum field theory,
with further complication of interaction among them in any realistic problems.
Even if this coupled system follows a universal change of oscillation
frequency given by a single function of time, one must be careful in
taking the time average. The simple average formula given above is valid
for time integration over time interval $\geq 2\pi / \omega $. Thus
the low frequency part must be treated separately if the corresponding
period is longer than the time interval considered. The infrared component
of the quantum field thus needs a special care.

To gain more insight it is instructive to explicitly
work out the following model of time variation:
\begin{eqnarray}
\omega ^{2}(t) = \omega _{0}^{2}\,\frac{1 + ae^{\kappa   t}}{1 + e^{\kappa  t}}
\,, \hspace{0.5cm} a > 0 \,, \hspace{0.5cm} \kappa  > 0 \,.
\end{eqnarray}
The oscillation frequency monotonically changes with a time scale
$1/\kappa $ like
\( \:
\omega(t) = \omega _{0} \longrightarrow \sqrt{a}\,\omega _{0} \equiv \omega
\: \)
as time evolves. It turns out that this model is analytically soluble at any
finite time \cite{landau-lifshitz qm}.
The solution to the Gelfand-Yaglom problem is expressed by the hypergeometric
function. For instance,
\begin{eqnarray}
\hspace*{-1cm}
\tilde{D}^{u}(t) =
\frac{1 + \epsilon }{2i\omega _{0}}
e^{i\omega _{0}t}\,
F(\,i\frac{\omega _{0}}{\kappa }(\sqrt{a} + 1) \,,
-\,i\frac{\omega _{0}}{\kappa }(\sqrt{a} - 1) \,,
1 + i\frac{2\omega_{0} }{\kappa } \,; -\,e^{\kappa t}) + ({\rm c.c.}) \,,
\end{eqnarray}
with
\( \:
F(\, \alpha \,, \beta \,, \gamma \,; z \,)
\: \)
the hypergeometic function.
For $t_{I} \rightarrow -\,\infty $ ,
\( \:
\epsilon = {\rm O \;}(e^{\kappa t_{I}}) \,,
\: \)
hence $\epsilon $ may be ignored.
Using the analytic continuation formula of the hypergeometric function,
one finds for the parameters in Eq.\ref{asymptotic sol} that
\begin{eqnarray}
&& \hspace*{-2cm}
a_{1} e^{i\delta _{1}} = \frac{1}{4\pi \kappa }\,e^{-\,i\omega _{0}t_{I}}\,
2^{i\frac{2\omega _{0}}{\kappa }(\sqrt{a} + 1)}\,
B(\,i\frac{\omega _{0}}{\kappa } \,, i\frac{\omega _{0}}{\kappa }\,\sqrt{a}\,)
\,B(\,\frac{1}{2} + i\frac{\omega _{0}}{\kappa } \,,
\frac{1}{2} + i\frac{\omega _{0}}{\kappa }\,\sqrt{a}\,)
\,, \nonumber \\
&& \hspace*{-2cm}
a_{2} e^{-\,i\delta _{2}} = \frac{1}{4\pi \kappa }\,e^{-\,i\omega _{0}t_{I}}\,
2^{-\,i\frac{2\omega _{0}}{\kappa }(\sqrt{a} - 1)}\,
B(\,i\frac{\omega _{0}}{\kappa } \,, -\,
i\frac{\omega _{0}}{\kappa }\,\sqrt{a}\,)
\,B(\,\frac{1}{2} + i\frac{\omega _{0}}{\kappa } \,,
\frac{1}{2} - i\frac{\omega _{0}}{\kappa }\,\sqrt{a}\,) \,,
\end{eqnarray}
with $B(\alpha \,, \beta )$ the Euler's beta function.
The crucial quantity $\Lambda = {\rm arctanh \;}(a_{2}/a_{1}) $
is then given by
\begin{eqnarray}
\tanh \Lambda = \frac{|\,\sinh (\,\pi (\omega - \omega _{0})/\kappa\,) \,|}
{\sinh (\,\pi (\omega + \omega _{0})/\kappa\,) } \,.
\end{eqnarray}
{}From this one obtains for the created particle number
\begin{eqnarray}
\langle \:N_{\omega } \: \rangle =
\frac{1}{2}\,
\frac{\cosh \frac{2\pi }{\kappa }(\omega  - \omega _{0}) - 1}
{\sinh \frac{2\pi \omega _{0}}{\kappa }\sinh \frac{2\pi \omega}{\kappa }} \,.
\end{eqnarray}
Let us take a limit of
\( \:
\omega _{0} \rightarrow \infty
\: \)
such that the final
\( \:
\omega = \sqrt{a}\,\omega _{0}
\: \)
is kept finite. Shortly afterwards we shall argue that this limit is
relevant to the de Sitter expansion.
The limit gives
\( \:
\tanh \Lambda \rightarrow e^{-\,2\pi \omega /\kappa } \,,
\: \)
and the produced number of particles
\( \:
\langle \:N_{\omega } \: \rangle \:\rightarrow  \:
\frac{1}{e^{4\pi \omega / \kappa } - 1} \,.
\: \)
This is exactly of Planckian form with the temperature:
\( \:
T = \frac{\kappa }{4\pi } \,.
\: \)

Let us now discuss an application of the analytic model just solved.
In the $a \rightarrow 0$ limit there exists an intermediate epoch
in which $e^{\kappa t} \gg 1$ but $a\,e^{\kappa t} \ll 1$, namely
\( \:
1/\kappa \ll t \ll -\,\ln a /\kappa \,.
\: \)
In this epoch
\( \:
\omega (t) \,\propto \, e^{-\,\kappa t/2} \,.
\: \)
This behavior may be considered as an analogue model of de Sitter inflation.
Suppose that all modes of quantum fields universally follow this exponential
decrease of frequency, as occurs for massless fields.
The parameter $\kappa $ is then interpretated as the Hubble rate
of expansion: $H = \kappa /2$.
(Strictly, modes of large physical wavelength
$\geq 1/H $ do not experience the exponential frequency decrease, and must
be treated separately.)
After the epoch of de Sitter phase the space expansion ends
and the universe settles to a mush slower phase of expansion. If this final
universe and also the initial one are approximated by the static universe,
one may consider the final produced particle being created by the de Sitter
expansion. Indeed, the resulting Planck distribution coincides with
that of Gibbons-Hawking \cite{gh}, including the temperature relation:
\( \:
T = \frac{H}{2\pi } \,.
\: \)
We find this interpretation worthy of further investigation, because
in the original approach of Gibbons and Hawking the thermal radiation
is observer dependent and its reality is not unambiguously established.
On the other hand, in our model the time averaged density matrix is nearly,
but not completely thermal. This and the infrared problem mentioned above
must further be examined.

\vspace{0.5cm}
In summary, we have given the general formula of the transition amplitude
for time varying harmonic oscillator, and in particular the wave function
of the evolved quantum ground state Eq.\ref{wave function} and its Wigner
function Eq.\ref{wigner function}. The resulting number of created particles
approaches a constant value if both the initial and the final oscillator
frequencies have well defined limits. We have also discussed how the
thermal behavior may emerge  from a quantum state via coarse graining.

\end{document}